\documentclass[a4paper,10pt]{article}

\usepackage[utf8]{inputenc}
\usepackage[T1]{fontenc}
\usepackage[frenchb,english]{babel}
\usepackage{textcomp}
\usepackage{amsmath}
\usepackage{amsfonts}
\usepackage{amssymb}
\usepackage{graphicx}
\usepackage{vmargin}
\usepackage[bookmarks=true,bookmarksnumbered=true,colorlinks=false,pdfborder={0 0 0}]{hyperref}
\usepackage{fancyhdr}
\usepackage{lastpage}
\usepackage{cases}
\usepackage{booktabs}

\setmarginsrb{25mm}{20mm}{25mm}{15mm}{12pt}{5mm}{0pt}{11mm}

\begin{document}


\pagestyle{fancy}
\chead{}
\lhead{\textit{B. Lenoir, 2013}}
\rhead{\textit{\thepage/\pageref{LastPage}}}
\cfoot{}
\renewcommand{\headrulewidth}{0pt}
\renewcommand{\footrulewidth}{0pt}



\title{Predicting the variance of a measurement with 1/f noise}
\author{Benjamin Lenoir\textsuperscript{a} \\ \small \textsuperscript{a} \textit{Onera -- The French Aerospace Lab, 29 avenue de la Division Leclerc, F-92322 Ch\^atillon, France} \\ \\ \small Published in \textit{Fluctuation and Noise Letters} 12:1 (2013) 1350006 \\ \small doi: \href{http://dx.doi.org/10.1142/S0219477513500065}{10.1142/S0219477513500065}}
\date{17 May 2013}
\maketitle



\begin{abstract}
Measurement devices always add noise to the signal of interest and it is necessary to evaluate the variance of the results. This article focuses on stationary random processes whose Power Spectrum Density is a power law of frequency. For flicker noise, behaving as $1/f$ and which is present in many different phenomena, the usual way to compute the variance leads to infinite values. This article proposes an alternative definition of the variance which takes into account the fact that measurement devises need to be calibrated. This new variance, which depends on the calibration duration, the measurement duration and the duration between the calibration and the measurement, allows avoiding infinite values when computing the variance of a measurement.

\paragraph{Keywords} 1/f noise; variance; precision; calibration; stationary stochastic process; power spectrum density.
\end{abstract}



\section{Introduction}\label{section:introduction}

This article deals with stationary random processes whose power spectrum density (PSD) $S(f)$ behaves as
\begin{equation}
 \forall f\in\mathbb{R}, \ S(f) = h |f|^\alpha
\end{equation}
with $\alpha\in]-3;+\infty[$. The condition on $\alpha$ allows covering ``flicker noise'' ($\alpha = -1$), which is the focus of this article, as well as white noise ($\alpha = 0$). The fact that flicker noise describes correctly experimental phenomena in many different fields has been verified~\cite{hooge1976noise,voss1979noise,mandelbrot1968fractional,voss1978noise}.

Because of the infinite value taken by $S(f)$ when $f \rightarrow 0$, discussions focus on the extend of the $1/f$ spectrum at low frequencies. However, the yet imperfectly understood mechanisms of the $1/f$ noise do not give reasons to expect a low-bound on the frequency range~\cite{flinn1968extent}. The problem becomes more acute when one tries to compute the variance of a measurement made with an instrument having this type of noise, since it leads to infinite values. One way to tackle this problem is to use a ``conditional spectrum''~\cite{mandelbrot1967some}. The idea is to take into account the fact that the physical phenomenon leading to the $1/f$ noise has been observed for a finite amount of time called $t^*$. If, during the measurement process, the signal is averaged over a period called $T$, the variance of the measured average scales as $\log(t^* / T)$ \cite{keshner1982noise,kleinpenning1988relation}.
This approach leads to a finite value for the variance. But it introduces the parameter~$t^*$, which may be completely arbitrary when one is interested by the average over the period $T$, and adds the unknown offset $\log(t^*)$.

This article is not concerned with the derivation of $|f|^\alpha$ noises from physical laws, nor with the idea of giving a lower bound to the frequency range of $1/f$ noise. Instead, the goal is to provide an efficient tool to compute the variance of a measurement made with flicker noise and more generally with a PSD leading to infinite variance due to its behavior close to zero. To do so, a new definition of the variance is introduced. It takes into account the measurement process, which is composed of the measurement itself but also of the calibration of the instrument. This quantity is first introduced for flicker noise. Some properties are given in this case and generalizations are made to other type of PSD.

\section{Introducing a Finite Measure of Variance with Flicker Noise}\label{section:introducing}

Let us consider a stationary stochastic process $n$ whose PSD is $S(f) = 1/|f|$. This noise is added to a deterministic signal called $s$. The aim of the measurement process considered here is to know the mean of the signal $s$ over a duration $T$ and to characterize the precision of this measurement. The usual way to proceed is to compute the variance $V(T)$ using the transfer function $h_T$ of the variance operator
\begin{equation}
 V(T) = 2 \int_0^\infty S(f) \left| H_T(f) \right|^2 df,
\end{equation}
where $H_T(f) = \text{sinc}(\pi T f)$ is the Fourier transform of $h_T$. $V(T)$ is equal to~$+ \infty$ because of the integrand being equivalent to $1/f$ when $f \rightarrow 0^+$.

\begin{figure}[ht]
\begin{center}
  \includegraphics[width=0.47\linewidth]{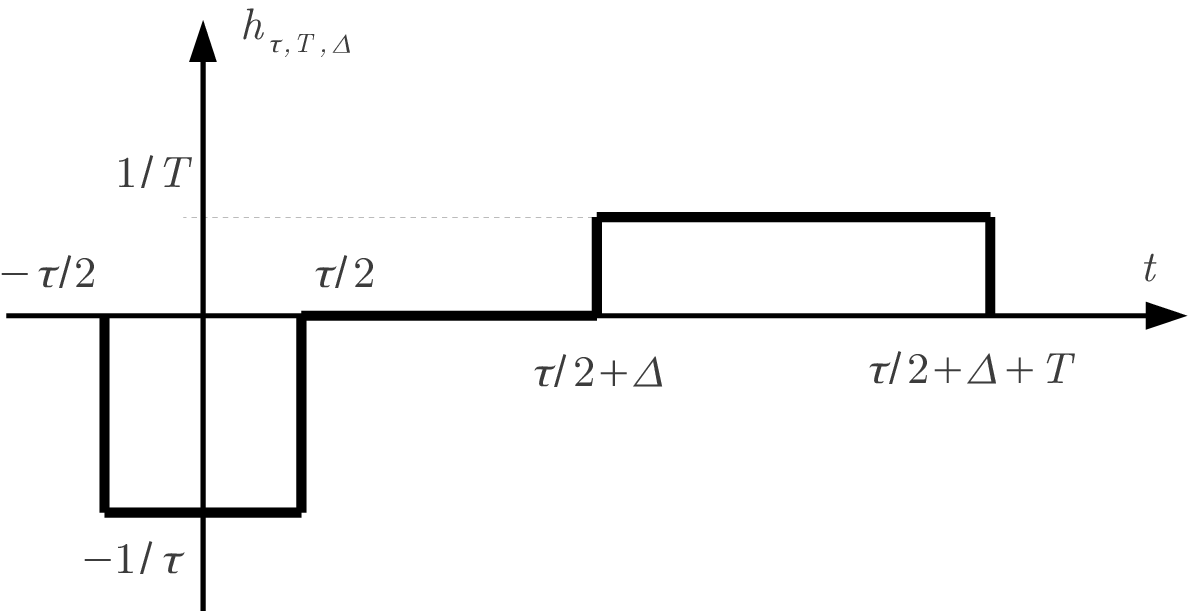}
  \hskip 20 pt
  \includegraphics[width=0.47\linewidth]{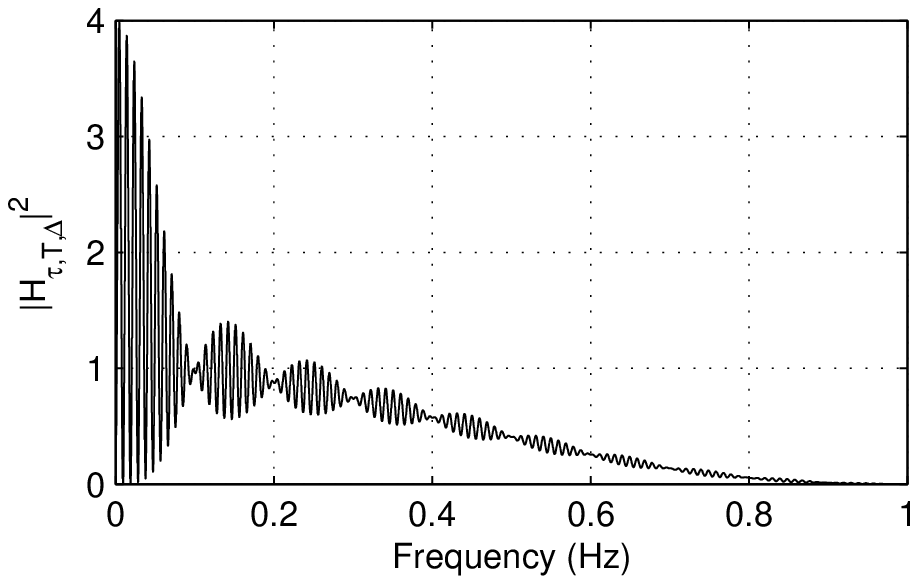}
  \caption{Transfer function of the variance $U$ defined in equation~\eqref{eq:new_var}. \textbf{Left:} In the temporal domain. \textbf{Right:} In the frequency domain with $T = 10$ s, $\tau = 1$ s and $\Delta = 100$ s.}
 \label{fig:transfer_function}
\end{center}
\end{figure}

The idea to get around this problem is inspired by the methodology used to make measurements. At some point in the measurement process, the instrument is calibrated, i.e. its bias is measured so as to be removed from the measurements of interest. The measurement equation of the instrument is
\begin{equation}
 m = s + b + n,
\end{equation}
where $m$ is the measurement, $s$ the signal of interest, $b$ the bias of the instrument and $n$ the noise. The calibration process consists in measuring subsequently the mean of $s$ and the mean of $-s$, each measurement lasting $ \tau/2$. Assuming that $s$ and $b$ are constant over $\tau$, it gives two quantities $m_{c1}$ and $m_{c2}$ such as
\begin{equation}
 b = \frac{m_{c1} + m_{c2}}{2} - \left<n\right>_\tau,
\end{equation}
where $\left<\cdot\right>_\tau$ is the mean over the duration $\tau$. As a result, when making a measurement of the signal~$s$ over a period $T$, the variance of the measurement depends on the measurement itself and the calibration of the bias, i.e. on the quantity
\begin{equation}
 -\frac{1}{\tau} \int_{-\tau/2}^{\tau/2} n(t) dt + \frac{1}{T} \int_{\tau/2+\Delta}^{\tau/2+\Delta+T} n(t) dt,
\end{equation}
with $\Delta$ the duration between the end of the calibration and the beginning of the measurement. This leads to consider the following new definition for the variance
\begin{equation}
 U(\tau,T,\Delta) = \mathbb{E}\left[ \left( - I_{-\tau/2}^{\tau} + I_{\tau/2 + \Delta}^{T} \right)^2 \right] = \mathbb{E}\left[ \left(n \ast h_{\tau,T,\Delta}\right)(t_0)^2 \right],
 \label{eq:new_var}
\end{equation}
where $I_a^b = \frac{1}{b} \int_a^{a+b} n(t) dt$, $\mathbb{E}$ is the expectation operator and $h_{\tau,T,\Delta}$ is defined in figure~\ref{fig:transfer_function}. $U(\tau,T,\Delta)$ does not depend on $t_0$, which is an arbitrary time, because $n$ is a stationary stochastic process.

\begin{figure}[ht]
\begin{center}
  \includegraphics[width=0.47\linewidth]{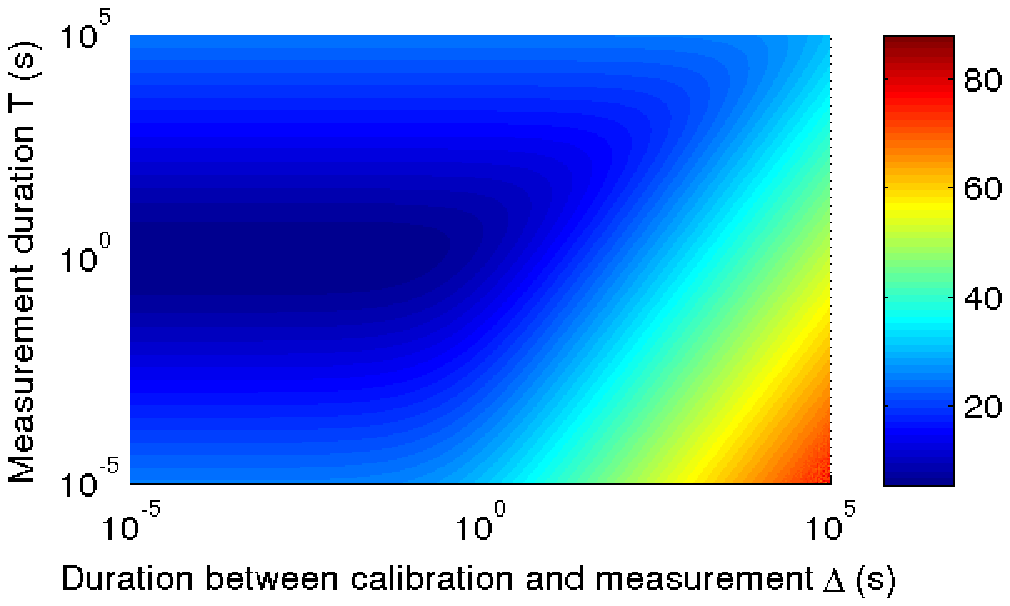}
  \hskip 20 pt
  \includegraphics[width=0.47\linewidth]{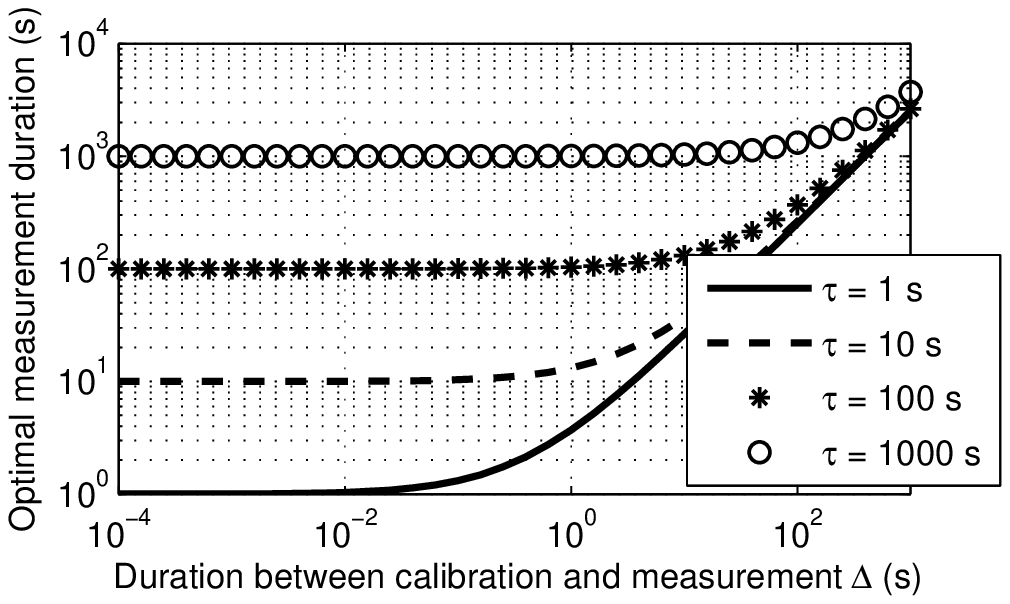}
  \caption{\textbf{Left:} Value of $U_{-1}$, defined by equation \eqref{eq:def_U-1}, as a function of the measurement duration~$T$, and the duration between the calibration and the measurement $\Delta$, for a calibration duration $\tau = 1$~s. \textbf{Right:} Value of the measurement duration $T$ for which $U_{-1}$ is minimum.}
 \label{fig:U-1}
\end{center}
\end{figure}

The transfer function of the variance~$U$, represented in figure~\ref{fig:transfer_function}, is equal in the Fourier domain to
\begin{equation}
 H_{\tau,T,\Delta}(f) = -\text{sinc}(\tau \pi f)^2 + \text{exp}\left[- i 2 \pi f \left(\Delta + \frac{T+\tau}{2}\right)\right] \text{sinc}(T \pi f)^2.
\end{equation}
This expression generalizes Allan variance~\cite{allan1966statistics} with uneven sizes for the samples and discontinuity between the samples. Since $|H_{\tau,T,\Delta}(f)|^2$ is equivalent to $f^2$ when $f \rightarrow 0^+$, the integral
\begin{equation}
 U_\alpha(\tau,T,\Delta) = 2 \int_0^\infty f^\alpha \left| H_{\tau,T,\Delta}(f) \right|^2 df,
 \label{eq:new_variance}
\end{equation}
is finite for $\alpha \in ]-3;1[$. For a flicker noise ($\alpha = -1$), computing the integral leads to the following expression:
\begin{align}
 U_{-1}(\tau,T,\Delta) = & -2 \ln(2 \pi \tau) - 2 \ln(2 \pi T) - \frac{2(\Delta + T)^2}{\tau T} \ln\left[2\pi(\Delta + T)\right] \nonumber \\
                         & - \frac{2(\Delta + \tau)^2}{\tau T} \ln\left[2\pi(\Delta + \tau)\right] + \frac{2\Delta^2}{\tau T} \ln\left[2\pi\Delta\right] \nonumber \\
                         & + \frac{2(\Delta + \tau + T)^2}{\tau T} \ln\left[2\pi(\Delta + \tau + T)\right].
 \label{eq:def_U-1}
\end{align}
Obviously, the following property is verified: $U_{-1}(\tau,T,\Delta) = U_{-1}(T,\tau,\Delta)$. It means that in term of variance the duration of the calibration and the measurement are equivalent. Figure \ref{fig:U-1} shows, for given $\tau$ and $\Delta$, the value of the integration time $T$ for which the variance reaches a minimum.

\section{Generalization}\label{section:results}

The method presented in the previous section can be applied to the other value of the coefficient $\alpha$. An exact result can be obtained for a white noise ($\alpha = 0$):
\begin{equation}
 U_{0}(\tau,T,\Delta) = \frac{1}{\tau} + \frac{1}{T}.
\end{equation}
In this case, the variance does not depend on the quantity $\Delta$. This is a characteristic of white noise because of the absence of correlation in the noise. Similarly, one can obtain analytically the value equation \eqref{eq:new_variance} for a brown noise ($\alpha = -2$):
\begin{equation}
 U_{-2}(\tau,T,\Delta) = 4 \pi^2 \left( \Delta + \frac{T}{3} + \frac{\tau}{3} \right).
\end{equation}
In this case, the standard deviation, which is the square root of the variance, scales as $\sqrt{\Delta}$. This is a characteristic of random walks which are described by brown noise.

\begin{figure}[ht]
\begin{center}
  \includegraphics[width=0.55\linewidth]{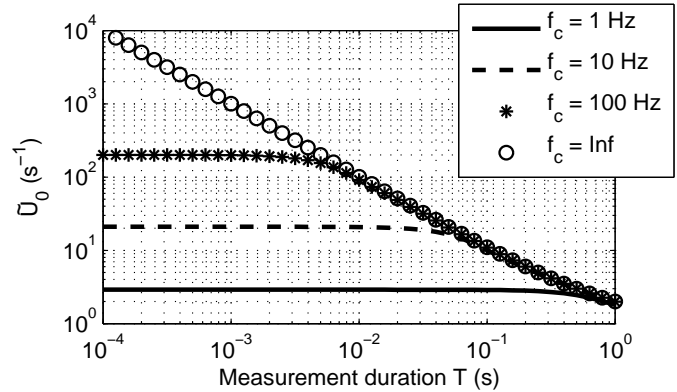}
 \caption{Value of $\tilde{U}_0$ as a function of the integration time $T$ for $\tau = 1$ s and $\Delta = 100$ s. The plots are indexed by the cut-off frequency of the measurement device.}
 \label{fig:U0}
\end{center}
\end{figure}

Finally, for values of $\alpha$ larger than $1$, it is necessary to take into account the cut-off frequency $f_c$ of the measurement device: the quantity $\tilde{U}_\alpha(\tau,T,\Delta,f_c)$ is the same as $U_\alpha$ but with an integration up to $f_c$ in equation~\eqref{eq:new_variance}. Therefore $U_\alpha(\tau,T,\Delta) = \tilde{U}_\alpha(\tau,T,\Delta,+\infty)$. This allows to solve the divergence of $U_\alpha$ for $\alpha\geq1$. Figure~\ref{fig:U0} shows the behavior of $\tilde{U}_\alpha$ for a white noise ($\alpha=0$) and for different values of $f_c$. The integration time $T$ for which the value of $\tilde{U}_0$ is different from the value of $U_0$ are those smaller than $1/f_c$.

\section{Conclusion}\label{section:conclusion}

This article aims at answering the following question while avoiding infinite values: what is the variance of a signal mean value measurement when the measurement noise is a flicker noise? To do so, a new definition of the variance was introduced. It was inspired by the usual measurement methodology which consists in calibrating the device before making measurements. This led to a variance which depends on three quantities : the calibration duration, the measurement duration and the duration between the calibration and the measurement. And it can be generalized for Power Spectrum Density of the form $|f|^\alpha$ with $\alpha > -3$.

\section*{Acknowledgments}\label{section:acknowledgments}

The author is grateful to CNES (Centre National d'\'Etudes Spatiales, France) for its financial support.




\end{document}